\documentstyle[11pt,moriond,epsfig]{article}

\def\be{\begin{equation}}
\def\ee{\end{equation}}
\def\bea{\begin{eqnarray}}
\def\eea{\end{eqnarray}}

\def\e{\mbox{e}}
\begin{document}
\vspace*{4cm}
\title{ULTRA-HIGH ENERGY COSMIC RAYS AND DIFFUSE 
PHOTON SPECTRUM}

\author{ P.G. TINYAKOV }

\address{Institute for Nuclear Research, \\ 
60th October Anniversary prospect 7a, 117312 Moscow, Russia}

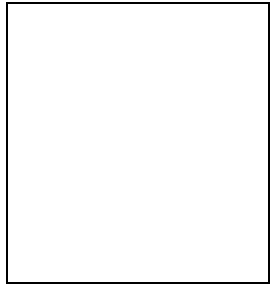
\maketitle\abstracts{ It is argued that if extragalactic magnetic
  fields are smaller than $2\times 10^{-12}$~G the flux of ultra-high
  energy photons of (a few)$\times
  10^{-1}$~eV$\,$cm$^{-2}\,$s$^{-1}\,$sr$^{-1}$ predicted in top-down
  models of UHE CR implies similar flux of diffuse photons in the
  energy range $10^{15}-10^{17}$~eV, which is close to the existing
  experimental limit.  }

In this talk I discuss possible relation between the flux of
Ultra-High Energy (UHE) photons and the diffuse photon flux in the
energy range $10^{15}-10^{17}$~eV, which occurs under certain
conditions. Namely, if extragalactic magnetic fields are sufficiently
small, a large fraction of energy carried by UHE photons is
transformed into $10^{15}-10^{17}$~eV photons, so that the energy
fluxes of these two components become comparable. In this way a flux
of $10^{15}-10^{17}$~eV photons can be produced which is close to the
existing experimental limit.

There are three main issues to be considered: propagation of the
diffuse photons, propagation of UHE photons and synchrotron radiation
in the inhomogeneous Galactic magnetic filed.  Let us start with
the diffuse photons. 

The detection of the diffuse photon flux at energies around $4\times
10^{14}$~eV has been reported by the Tian-Shan experiment
\cite{Tian-Shan} which gives the value $I_{\gamma}(E>4.0\times
10^{14}\,\mbox{eV}) = (3.4\pm 1.2)\times10^{-13}$~cm$^{-2}$s$^{-1}$.
More recently, the bounds have been obtained at higher energies, the
most stringent ones coming from EAS-TOP and CASA-MIA. These bounds are
$2.5\times10^{-14}$~cm$^{-2}$s$^{-1}$ at $E\sim 10^{15}$~eV
\cite{EAS-TOP,CASA-MIA} and $5.4\times10^{-17}$~cm$^{-2}$s$^{-1}$ at
$E\sim 2.2\times 10^{16}$~eV \cite{CASA-MIA}.

The sources of diffuse photons include, in particular, electromagnetic
cascades and secondary pions produced during the propagation of UHE CR
through intergalactic space. Regardless of the particular model the
spectrum of diffuse photons is expected to have a dip at energies
$10^{15}-10^{17}$~eV due to the e$^+$e$^-$--pair production on cosmic
microwave background \cite{dip}.  The cross section of the latter
reaction reaches maximum of 0.2~barn near the threshold at $3\times
10^{14}$~eV, so that the attenuation length of photons in this region
is as small as $\sim 20$ kpc. Thus, a large flux of photons in this
energy range can only be produced by nearby sources, while the absence
of the dip in the spectrum would require that these sources contribute
a large fraction of the total flux of diffuse photons.

One of the mechanisms which allows to produce the high energy photons
in our Galaxy makes use of the synchrotron radiation of the electron
component of UHE CR in the Galactic magnetic field. The UHE electrons
are generically produced in the halo models of UHE CR in the course of
the fragmentation process and rapidly convert their energy into
synchrotron radiation\cite{Blasi}. Another possibility which is not
related to halo models but requires large fraction of photons in UHE
CR and small extragalactic magnetic fields is that UHE electrons are
generated during the cascade propagation of UHE photons \cite{DT2}. In
this paper we concentrate on this latter possibility.

To make the discussion quantitative consider the equations which
describe the electromagnetic cascade in the absence of magnetic field
(the details can be found , e.g., in ref.\cite{Lee}). The main
reactions driving the cascade are $e^+e^-$ pair production (PP) on the
radio background, $\gamma\gamma_b \to e^+e^-$, double pair production
(DPP), $\gamma\gamma_b \to e^+e^-e^+e^-$, and inverse Compton
scattering (ICS), $e\gamma_b \to e \gamma$. If one neglects secondary
particles and energy losses then PP and DPP lead to the conversion of
photons to electrons with the rates $a_{PP}$ and $a_{DPP}$,
respectively, while ICS converts electrons back to photons with the
rate $b$. At the distance $R$ from the source the electron to photon
ratio is given by
\begin{equation}
{n_e\over n_{\gamma}} = {a\e^{R(a+b)}-C\over b\e^{R(a+b)}+C}\;, 
\label{ne-over-ng}
\end{equation}
where $a\equiv a_{PP} + a_{DPP}$ and $C$ is an integration constant
whose value is irrelevant far from the source. Since the observed 
fluxes are saturated by large distances, the ratio of electron to
photon flux is 
\begin{equation}
{F_e \over F_{\gamma}} \sim {a\over b}.  
\label{F/Fnaive}
\end{equation}
The latter ratio depends on energy. Numerically, 
\begin{eqnarray}\label{Fe/Fg22}
{F_e\over F_{\gamma}} &\sim &2 \quad\quad \mbox{at}\quad 
E= 10^{22}\;\mbox{eV} ,\\
{F_e\over F_{\gamma}} &\sim& 10 \quad\quad \mbox{at}\quad 
E= 10^{23}\;\mbox{eV}.
\label{Fe/Fg23}
\end{eqnarray}
Thus, the flux of UHE photons implies at least as large flux of UHE
electrons. The extragalactic magnetic field can substantially reduce
this ratio.  Estimates show that extragalactic magnetic fields smaller
than $2\times 10^{-12}$~G do not spoil our argument.

Finally, consider the synchrotron radiation of UHE electrons in the
Galactic magnetic field.  An ultra-relativistic particle of energy $E$
moving in the magnetic field $B$ emits radiation at the characteristic
frequency
\begin{equation}
\omega_c = {3\sqrt{\alpha} B\over 2 m_e^3} E^2 
= 6.7\times 10^{14} \left({E\over 10^{20}\,\mbox{eV}}\right)^2
\left({B\over 10^{-6}\,\mbox{G} } \right) \,\mbox{eV}. 
\label{omega-sync}
\end{equation}
As a result of this process, the particle looses energy at
the rate 
\begin{equation}
{dE\over dx} = - {2\alpha^2B^2\over 3m^4} E^2.
\label{energy-loss-eq}
\end{equation}
In the magnetic field of the form 
\begin{equation}
B(x) = B_0\exp(x/x_0),
\label{B}
\end{equation}
where $x_0\sim 4$~kpc \cite{Stanev}, the dominant radiation frequency
changes with particle energy according to the following equation,
\[
\omega_c(E) = {9 E_0^{3/2}\over 2 m_e \sqrt{3\alpha x_0}} f(E/E_0),
\]
where $f(y) = y^{3/2}(1-y)^{1/2}$ and $E_0$ is the initial energy of
the electron. Thus, most part of the electron energy is emitted at
frequencies close to
\begin{equation}
\omega_{max} = \omega_c(3E_0/4) = 
{27 E_0^{3/2}\over 32 m_e \sqrt{\alpha x_0}}
= 0.8\times 10^{15}\left( {E_0\over 10^{22}\,\mbox{eV} }\right)^{3/2}
\left( {x_0\over 4\;\mbox{kpc}} \right)^{-1/2} \mbox{eV}.
\label{omegamax}
\end{equation}
According to eqs.(\ref{omega-sync}) and (\ref{omegamax}), at
$E=10^{22}$~eV the electron looses most part of its energy in the
region where the magnetic field is $B \sim 10^{-10}$~G. 

Finally, let us estimate the flux of synchrotron photons assuming the
flux of UHE photons typical for top-down scenarios, $(\mbox{a
  few})\times 10^{-1}$~eV$\,$cm$^{-2}$s$^{-1}$sr$^{-1}$ at energies
$\sim 10^{22}-10^{23}$~eV. Eq.(\ref{Fe/Fg22}) implies that outside the
Galactic magnetic field there is at least as large flux of UHE
electrons which transfer their energy to high energy photons in the
Galactic magnetic field. Since the synchrotron spectrum has
$\delta\omega\sim \omega$, the energy conservation implies that the
flux of synchrotron photons is approximately the same as the flux of
UHE electrons, which is larger by the factor $F_e/F_{\gamma}$ than the
flux of UHE photons. One therefore obtains the diffuse photon flux in
the range $10^{15}-10^{17}$~eV of the order of $\sim
1$~eV$\,$cm$^{-2}$s$^{-1}$sr$^{-1}$.

To summarise, we have shown that the expected dip in the photon
spectrum at energies $10^{15}-10^{17}$~eV may be filled by the
synchrotron radiation of UHE CR in the Galactic magnetic field. The
mechanism we propose requires small extragalactic magnetic fields and
large fraction of photons in UHE CR at energies $10^{22}-10^{23}$~eV.
The last requirement is satisfied in the top-down models of UHE CR.
The flux of $10^{15}-10^{17}$~eV photons produced by our mechanism in
these models is close to the present experimental limit.  Thus, it is
important to improve the sensitivity of the experiments in the energy
range $10^{15}-10^{17}$~eV. The detection of the diffuse photon flux
at the level of $\sim 10^{-1}$~eV$\,$cm$^{-2}$s$^{-1}$sr$^{-1}$ would
strongly suggest that UHE CR are produced by a top-down mechanism.  On
the contrary, if the photon flux in the region of the dip is smaller
than $\sim 10^{-3}$~eV$\,$cm$^{-2}$s$^{-1}$sr$^{-1}$ and, at the same
time, UHE CR have a large fraction of photons, the EGMF must be larger
than $2\times 10^{-12}$~G. The detailed calculation of the high energy
photon spectrum in various models of UHE CR which takes into account
synchrotron radiation in the Galactic magnetic field will be published
elsewhere.

\section*{Acknowledgements}

The author acknowledges the hospitality of the organisers of the ``XI
Rencontres de Blois'' and the financial support of the Russian
Foundation for Basic Research, travel grant 99-02-78141.

\end{document}